\documentclass{sig-alternate-05-2015}

\usepackage{mathpartir}
\usepackage{verbatim}
\usepackage{alltt}
\usepackage{url}
\usepackage{breakurl} 
\usepackage[breaklinks]{hyperref}
\usepackage{stmaryrd}
\usepackage[usenames,dvipsnames]{color}
\usepackage{etoolbox}
\usepackage{balance}
\usepackage{lipsum}
\usepackage{caption}
\usepackage{algorithm}
\usepackage[noend]{algpseudocode}
\usepackage{graphicx}
\usepackage{comment}
\usepackage{bbm, dsfont}
\usepackage{mathrsfs}
\usepackage{xcolor}
\usepackage{esvect}
\usepackage{xspace}
\usepackage{paralist}
\usepackage{tikz}
\usepackage{pgfplots}
\usepackage{subfigure}
\usepackage{xfrac}
\usepackage{adjustbox}
\usepackage{fancyvrb}

\usepackage{amsthm}
\usepackage{pifont}

\newcommand{\system}{{\sc Hunter}\xspace}
\newcommand{\proc}{{\sc CodeReuse}\xspace}
\newcommand{\desc}{\mathcal{D}}
\newcommand{\sig}{\mathcal{S}}
\newcommand{\tests}{\mathcal{T}}
\newcommand{\map}{\mathcal{M}}
\newcommand{\adapter}{\mathcal{R}}
\newcommand{\adaptee}{\mathcal{E}}
\newcommand{\ssix}{{\sc S$^6$}\xspace}

\definecolor{bblue}{HTML}{0064FF}
\definecolor{rred}{HTML}{C0504D}
\definecolor{ggreen}{HTML}{9BBB59}
\definecolor{ppurple}{HTML}{9F4C7C}

\newcommand{\cmark}{\textcolor{ggreen}{\ding{51}}}
\newcommand{\xmark}{\textcolor{rred}{\ding{55}}}

\newenvironment{myverb}{%
 \VerbatimEnvironment
 \begin{adjustbox}{max width=\linewidth}
 \begin{BVerbatim}
  }{
  \end{BVerbatim}
 \end{adjustbox}
}

\newtheorem{theorem}{Theorem}[section]
\newtheorem{definition}[theorem]{Definition}
\newtheorem{example}{Example}

\newcommand*{\comments}{} 
\ifdefined\comments
\newcommand{\yu}[1]{\textcolor{blue}{\textbf{YU:} #1}}
\newcommand{\yuepeng}[1]{\textcolor{cyan}{\textbf{YUEPENG:} #1}}
\newcommand{\arati}[1]{\textcolor{green}{\textbf{ARATI:} #1}}
\newcommand{\ruben}[1]{\textcolor{magenta}{\textbf{RUBEN:} #1}}
\newcommand{\isil}[1]{\textcolor{red}{\textbf{I\c{S}IL:} #1}}
\newcommand{\todo}[1]{\textcolor{red}{#1}}
\else
\newcommand{\yu}[1]{}
\newcommand{\yuepeng}[1]{}
\newcommand{\arati}[1]{}
\newcommand{\ruben}[1]{}
\newcommand{\isil}[1]{}
\newcommand{\todo}[1]{}
\fi

\newcommand{\irule}[2]%
   {\mkern-2mu\displaystyle\frac{#1}{\vphantom{,}#2}\mkern-2mu}

\makeatletter
\def\@copyrightspace{\relax}
\makeatother

\begin{document}



\title{Type-Directed Code Reuse \\ using Integer Linear Programming}

\numberofauthors{6}
\author{
\parbox{5cm}{\centering
Yuepeng Wang\\
\affaddr{University of Texas at Austin}\\
\email{ypwang@cs.utexas.edu}\\}
\and
\parbox{5cm}{\centering
Yu Feng\\
\affaddr{University of Texas at Austin}\\
\email{yufeng@cs.utexas.edu}\\}
\and
\parbox{5cm}{\centering
Ruben Martins\\
\affaddr{University of Texas at Austin}\\
\email{rmartins@cs.utexas.edu}\\}
\and
\parbox{5cm}{\centering
Arati Kaushik\\
\affaddr{University of Texas at Austin}\\
\email{arati@cs.utexas.edu}\\}
\and
\parbox{5cm}{\centering
Isil Dillig\\
\affaddr{University of Texas at Austin}\\
\email{isil@cs.utexas.edu}\\}
\and
\parbox{5cm}{\centering
Steven P. Reiss\\
\affaddr{Brown University}\\
\email{spr@cs.brown.edu}\\}
}

\maketitle

\begin{abstract}
In many common scenarios, programmers need to implement functionality that is already provided by some third party library. 
This paper presents a tool called \system that facilitates code reuse by finding relevant methods in large code bases and \emph{automatically synthesizing} any necessary wrapper code. The key technical idea underlying our approach is to use \emph{types} to both improve  search results and  guide synthesis. Specifically, our method computes \emph{similarity metrics} between types and uses this information to solve an  \emph{integer linear programming (ILP)} problem in which the objective is to minimize the cost of synthesis.  We have implemented \system as an Eclipse plug-in and evaluate it by (a) comparing it against \ssix, a state-of-the-art code reuse tool, and (b) performing a user study.   Our evaluation shows that \system compares favorably with \ssix and significantly increases programmer productivity.

\end{abstract}

\section{Introduction}
\label{sec:intro}

In many common scenarios, programmers need to implement functionality that is already provided by some third party library. In such cases, accepted wisdom dictates that it is preferable to reuse existing code rather than reimplementing the desired functionality from scratch. Indeed, there are at least two benefits to software reuse: First, 
it increases programmer  productivity, allowing developers to focus on more creative tasks. Second, implementations provided by existing APIs are typically well-tested; hence, code reuse decreases the likelihood that the implementation is buggy. 


Unfortunately, there is often a wide gap between the adage ``software reuse is good" and the reality of software development: In many cases,  programmers end up reimplementing functionality that already exists in a library. This duplicated effort is sometimes inadvertent because programmers may not know whether the desired functionality exists or which third-party library provides it. In other cases, the exposed interface may not quite fit the developer's needs, tempting programmers to reimplement the desired functionality. 

This paper seeks to address both of these problems by presenting a tool called \system for automatically reusing existing code. Given a few \emph{test cases} and a \emph{search query} provided by the user, \system searches existing code bases for candidate adaptee methods  and \emph{automatically synthesizes} any necessary wrapper code.  The key technical idea underlying our approach is to use \emph{types} to both improve  search results and  guide synthesis. Specifically, our method computes \emph{similarity metrics} between types and uses this information to solve an  \emph{integer linear programming (ILP)} problem in which the objective is to minimize the cost of synthesis.

\begin{figure*}[!t]
\begin{center}
\includegraphics[scale=0.27]{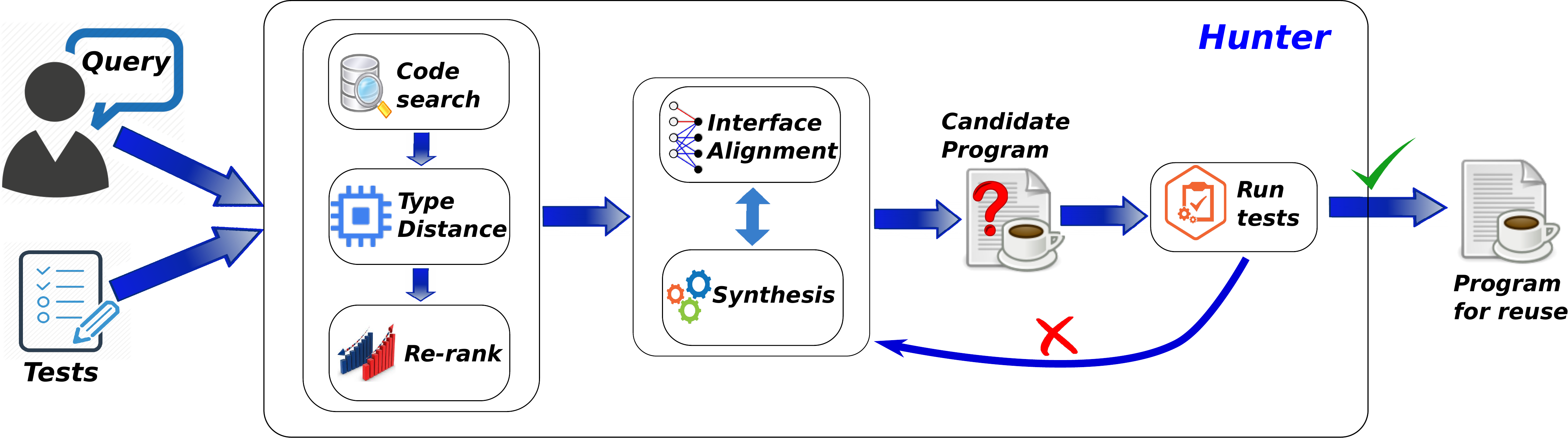}
\end{center}
\vspace{-0.1in}
\caption{Workflow of the \system tool}\label{fig:hunter}
\end{figure*}

The high-level architecture of the \system tool is shown in Figure~\ref{fig:hunter}. Our approach consists of three components, namely \emph{code search}, \emph{interface alignment}, and \emph{synthesis}. Given a candidate adaptee identified through code search, \emph{interface alignment} refers to the problem of finding a suitable mapping from parameters in the adaptee to those in the adapter. Given a solution to the interface alignment problem, \system uses existing synthesis tools to generate the desired wrapper code. If the synthesized program does not pass the provided test cases, \system \ backtracks by finding another candidate adaptee or a different solution to the interface alignment problem.


The key technical contribution of this paper is a type-directed approach for solving the interface alignment problem. Specifically, \system computes  \emph{distance metrics} between types in order to estimate how likely it is that an object of type $\tau_1$ can be used in the place of another object of type $\tau_2$. Specifically, the larger the distance $d$, the less likely it is that $\tau_1$ can be ``converted" to $\tau_2$. Given such a \emph{type distance matrix}, \system solves an optimization problem in which the goal is to minimize the cost of converting between the type signature of the adaptee and that of the desired method. Specifically, we formulate interface alignment as an integer linear programming problem that is parametrized by the type distance matrix. 

We have implemented \system as an Eclipse plug-in and evaluate it by performing three sets of experiments. First, we compare \system against \ssix,   a state-of-the-art code reuse tool, and show that \system can reuse more code compared to \ssix. Second, we perform a user study and show that \system  enables users to finish programming tasks quicker and with fewer bugs. Third, we  show that the type distances computed by \system can be gainfully used to improve the results of existing code search tools.
To summarize, this paper makes the following contributions:

\begin{itemize}
\vspace{-0.05in}
\item We describe a novel code reuse tool, \system, that combines code search with program synthesis. \system is available as an Eclipse plug-in and can automatically generate adapter methods that reuse existing Java methods in massive open source repositories.
\item We propose a technique for measuring similarity between arbitrary Java types and leverage it to re-rank search results obtained from existing code search tools. 
\item We define the \emph{interface alignment problem} for matching a pair of method signatures and show how it can be solved using integer linear programming (ILP).
\item We perform an extensive evaluation of \system by comparing it against a state-of-the-art code reuse tool (\ssix), and  conducting a user study. Our evaluation shows that \system compares favorably with \ssix and significantly increases programmer productivity.
\end{itemize}

\section{Overview}\label{sec:overview}
In this section, we give a high-level overview of the \system system through a simple motivating example.

\begin{figure}
\begin{myverb}
@Test public void test() {
    Vector<MyPoint> v1 = new Vector<MyPoint>();
    v1.add(new MyPoint(0, 0)); v1.add(new MyPoint(1, 1));
    v1.add(new MyPoint(2, 1)); v1.add(new MyPoint(3, 2));
    v1.add(new MyPoint(4, 2)); v1.add(new MyPoint(5, 3));
    Bresenham.drawLine(new MyPoint(5, 3), res);
    assertEquals(v1, res);
}
\end{myverb}
\vspace{-0.2in}
\caption{Test cases for the desired {\tt drawLine} method}\label{fig:test}
\end{figure}

\subsection{Motivating Example}
\label{sec:example}

Consider a user, Alice, who needs to draw a straight line from  the origin to a point. Specifically, Alice would like to implement in Java the following \verb+drawLine+ method:

{\small
\begin{verbatim}
void drawLine(MyPoint pt, Vector<MyPoint> res)
\end{verbatim}
}

Here, \verb+pt+ is the point specified by the user, and \verb+res+ is a vector of points on the raster that should be selected to approximate a straight line between the origin and \verb+pt+.  Alice knows about  Bresenham's line drawing algorithm that could be used to implement this functionality, but she does not know exactly how it works. While she would like to reuse an existing implementation of Bresenham's algorithm, she cannot find an implementation that quite fits her needs. 

The \system tool could help a user like Alice by finding an existing implementation of Bresenham's algorithm and automatically synthesizing all the wrapper code needed to integrate it into the desired \verb+drawLine+ interface. In order to use \system, Alice needs to provide the method signature shown above as well as a a brief natural language description, such as {\it ``Draw a line between the origin and specified point using Bresenham's algorithm''}. Alice also needs to provide a small test suite (e.g., the one shown in Figure~\ref{fig:test}) that \system can use to validate the synthesized code.

\system first  queries a code search engine using Alice's description and retrieves the top $k$ results. To simplify the example, suppose that the search engine yields a single function with the following signature:

{\small
\begin{verbatim}
Point[] bresenham(int x0, int y0, int x1, int y1)
\end{verbatim}
}

Specifically, this function returns an array of {\tt Point}s to approximate a line starting from $(x0, y0)$ to $(x1, y1)$. Note that there are several differences between Alice's desired \verb+drawLine+ interface and the existing \verb+bresenham+ function:

\begin{itemize}
\item Alice's interface uses \verb+MyPoint+ to represent the user-specified  point, while the existing function uses two integers (namely, \verb+x1+ and \verb+y1+) to represent the end point.
\item Alice's interface assumes the origin as a starting point, whereas \verb+bresenham+   takes \verb+x0+ and \verb+y0+ as input.
\item The existing \verb+bresenham+ function returns an array of \verb+Point+'s, whereas Alice's interface ``returns" the  line by storing the result in  \verb+res+. Furthermore, Alice represents  points using a custom type called \verb+MyPoint+, whereas \verb+bresenham+ uses a different type called \verb+Point+.
\end{itemize}

Despite these significant differences, \system is able to to automatically generate the following implementation of Alice's \verb+drawLine+ interface:
\vspace*{0.1in}

{\small
\begin{myverb}
void drawLine (MyPoint pt, Vector<MyPoint> res) {
    int v1 = pt.getX(); 
    int v2 = pt.getY();
    Point[] v3 = external.bresenham(0, 0, v1, v2);
    for (Point v4 : v3) {
        int v5 = v4.getX();
        int v6 = v4.getY();
        MyPoint v7 = new MyPoint(v5, v6);
        res.add(v7);
    }
}
\end{myverb}
\vspace{1mm}
}

Observe that the code generated by \system first deconstructs the \verb+MyPoint+ object \verb+pt+ into a pair of integers by invoking the appropriate getter methods.  Also note that \system can supply default values for \verb+x0+ and \verb+y0+ even though there are no corresponding parameters in the \verb+drawLine+ interface. Finally, after invoking the existing \verb+bresenham+ method, \system can synthesize code to convert the array of points into a vector of \verb+MyPoint+s.


\subsection{System Overview}

The high-level structure of our code reuse algorithm is presented in Algorithm~\ref{algo:hunter}. The \proc procedure takes as input a desired signature $\sig$, a natural language description $\desc$, and a test suite $\tests$. The return value of \proc is an implementation $\adapter$ of $\sig$ that passes all test cases if one exists, and $\bot$ otherwise.


\algdef{SE}[DOWHILE]{Do}{doWhile}{\algorithmicdo}[1]{\algorithmicwhile\ #1}%

\begin{algorithm}[t]
\caption{Code Reuse Algorithm}
\label{algo:hunter}
\begin{algorithmic}[1]
\vspace{0.05in}
\Procedure{\proc}{$\sig$, $\desc$, $\tests$}
\vspace{0.05in}
\State {\rm \bf Input:} Signature $\sig$ of desired method, natural \\
\ \ \ \ \ \ \ \ \ \ \ \ \ \ \ \ language description $\desc$, and tests $\tests$
\State {\rm \bf Output:} adapter $\adapter$ or failure $\bot$
\vspace{0.05in}

\State [$\adaptee_0$] := {\sc CodeSearch}($\sig, \desc$) \vspace{0.05in}
\State $\Lambda$ \ \ := {\sc ComputeTypeDistance}($\sig, [\adaptee_0]$) \vspace{0.05in}
\State [$\adaptee$] := {\sc Rerank}($[\adaptee_0], \sig, \Lambda$) \vspace{0.05in}
\ForAll{$\adaptee \in [\adaptee]$} \vspace{0.05in}
    \Do
    \State $\map$ := {\sc GetBestAlign}($\sig$, $\adaptee$, $\Lambda$) \vspace{0.05in}
     \State  $\adapter$ := {\sc AdapterGen}($\map, \adaptee, \sig$) \vspace{0.05in}
      \If{{\sc RunTests}($\adapter, \tests$)}
            \State \Return ($\adapter$, $\adaptee$)
       \EndIf
    \doWhile{$\map \neq \emptyset$}
\EndFor
\State \Return $\bot$

\EndProcedure
\end{algorithmic}
\end{algorithm}

As shown in Algorithm~\ref{algo:hunter}, our technique first uses a code search engine 
to retrieve a ranked list $[\adaptee_0]$ of relevant methods that could serve as possible adaptees. Next, we invoke the {\sc ComputeTypeDistance} procedure to compute a distance between each pair of types $(\tau, \tau')$, where $\tau$ is a type used in signature $\sig$ and $\tau'$ is a type used in the signature of some $\adaptee_0 \in [\adaptee_0]$. The return value $\Lambda$ of {\sc ComputeTypeDistance} is a matrix that maps each pair of types $(\tau, \tau')$ to a distance. The larger the distance, the less similar $\tau$ is to $\tau'$ and the less likely it is that an  argument of type $\tau$ will be mapped to an  argument of type $\tau'$. 

Once we compute the \emph{type distance matrix} $\Lambda$, \system re-ranks the original search results 
%
%
and tries to synthesize wrapper code for each candidate adaptee $\adaptee \in [\adaptee]$. 
Towards this goal, we first solve the interface alignment problem using integer linear programming.
%
Specifically, the call to the {\sc GetBestAlign} procedure at line 10 returns a lowest cost mapping between the parameters in the adapter and those in the adaptee. For example, the alignment between \verb+drawLine+ and \verb+bresenham+ from Section~\ref{sec:example} is given by the  mapping:

{\small
\begin{verbatim}
x1: int       ->   pt: MyPoint
y1: int       ->   pt: MyPoint
v3: Point[]   ->  res: Vector<MyPoint>
\end{verbatim}
}

Intuitively, this alignment minimizes the objective function in our ILP problem according to the costs given by $\Lambda$. 
Note that every parameter in the adapter must be mapped to some parameter in the adaptee because an implementation of $\adapter$ that does not use one of $\adapter$'s arguments is extremely unlikely to be correct. On the other hand, we do \emph{not} require that every parameter in the adaptee to be mapped to a parameter in the adapter, since  the implementation of the adaptee may be more generic. 


Given a candidate alignment $\mathcal{M}$, our algorithm invokes a procedure called {\sc AdapterGen} to synthesize wrapper code based on this alignment. For instance, for our running example, {\sc AdapterGen} synthesizes code to ``convert" the \verb+pt+ object of type \verb+MyPoint+ to an integer \verb+x1+ of type \verb+int+. Since there are several existing tools~\cite{jungloid,insynth,codehint,sypet} for type-directed code synthesis, our {\sc AdapterGen} procedure  uses existing synthesis techniques  to  generate a code snippet for ``converting" an object of type $\tau$ to an  object of type $\tau'$.

Once \system generates wrapper code for a given candidate alignment, we run the provided test cases to check if any test fails. If so, we backtrack and ask the ILP solver for the next best alignment, if one exists. If we exhaust all possible alignments for a given candidate $\adaptee$, we also backtrack and try the next best candidate adaptee. The \proc algorithm terminates when we  find an implementation that passes all test cases or we run out of possible adaptees.

\section{Type Distance}
\label{sec:distance}

In this section, we describe our technique for computing distances between a pair of types. As mentioned earlier, our approach uses type distances to both re-rank search results and compute a best alignment between the signature of the desired method and that of the candidate adaptee.

\subsection{Multiset Representation of Types}

In order to compute a distance metric between a pair of types, our approach represents each Java type using a \emph{multiset (bag) representation}. Given a type $\tau$, we use the notation $\psi(\tau)$ to denote $\tau$'s multi-set representation. Intuitively, $\psi(\tau)$ represents $\tau$ as a bag of \emph{features}. In this context, a feature is either a Java type or a boolean attribute, such as \emph{numeric}, \emph{collection} etc. As we will see in the next subsection, our approach computes a distance between two types $\tau$ and $\tau'$ using the multi-set representations $\psi(\tau)$ and $\psi(\tau')$.

%
%

We define the multiset representation $\psi(\tau)$ of each type $\tau$ using the inference rules shown in Figure~\ref{fig:typerules}.
The first rule, {\it Primitive I}, states that the multi-set representation of a numeric primitive type $\tau$, such as {\tt int} or {\tt double}, includes both the type name as well as the attribute {\tt numeric}.~\footnote{Boxed types of primitives are considered as their corresponding primitive types.}  On the other hand, if a primitive type is not numeric, then {\it Primitive II} states that its multiset representation consists of only the type name. For instance,  we have $\psi({\tt int})$ $=$ $ {\tt \{int, numeric\}}$, and $\psi({\tt String}) = {\tt \{String\}}$.


To capture the different collection types, we define {\tt cate\-gory}$(\tau)$ as a function that returns the collection type of $\tau$.
Specifically, we classify  Java built-in collections into four different categories, namely {\tt Vector}, {\tt List}, {\tt Set}, and {\tt Map}. For instance, we have {\tt category(ArrayList)} =  {\tt List} and {\tt category(TreeMap)} = {\tt Map}. Hence, according to the \emph{Collection} rule, the multi-set representation for a collection $\tau$ is given by $\{ {\tt category}(\tau), {\tt collection} \}$. 


%

\begin{figure}[!t]
\begin{center}
\[
\begin{array}{cr}

\irule
{\emph{isPrimitive}(\tau), \emph{isNumeric}(\tau) }
{\psi(\tau) = \{\tau, {\tt numeric}\} } & ({\rm Primitive~I}) \\ \ \\

\irule
{\emph{isPrimitive}(\tau), \emph{$\neg$isNumeric}(\tau) }
{\psi(\tau) = \{\tau\} } & ({\rm Primitive~II}) \\ \ \\

\irule
{\emph{isCollection}(\tau) }
{\psi(\tau) = \{{\tt category}(\tau), {\tt collection}\} } & ({\rm Collection}) \\ \ \\

\irule
{\emph{isArray}(\tau), \tau = \tau'[ ] }
{\psi(\tau) = \psi(\tau') \cup \{{\tt Array}, {\tt collection}\} } & ({\rm Array}) \\ \ \\


\irule
{\emph{isWildcard}(\tau) }
{\psi(\tau) = \emptyset } & ({\rm Wildcard}) \\ \ \\

\irule
{\emph{isTypeParam}(\tau) }
{\psi(\tau) = \emptyset } & ({\rm TypeParam}) \\ \ \\

\irule
{
\begin{array}{c}
\emph{isParameterized}(\tau) \\ \tau = t \langle \tau_1, \ldots, \tau_n \rangle 
\end{array} }
{\psi(\tau) = \psi(t) \cup \psi(\tau_1) \cup \cdots \cup \psi(\tau_n) } & ({\rm Parameterized}) \\ \ \\

\irule
{\begin{array}{c}
    \emph{isRef}(\tau),  \tau = t \{ L \}  \\
\end{array}}
{\psi(\tau) = \{t\} \cup \psi(L)} & ({\rm RefType}) \\ \ \\

\irule
{ 
L = \emptyset }
{\psi(L) = \emptyset } & ({\rm FieldList~I}) \\ \ \\

\irule
{
    L = \tau_1 ; L_1,  
    \emph{isRef}(\tau_1)
}
{\psi(L) = \{ \tau_1 \} \cup \psi(L_1)} & ({\rm FieldList~II}) \\ \ \\

\irule
{ 
    L = \tau_1 ; L_1,   
    \emph{$\neg$isRef}(\tau_1)
}
{\psi(L) = \psi(\tau_1) \cup \psi(L_1)} & ({\rm FieldList~III})

\end{array}
\]
\end{center}
\caption{Multiset representation of Java types}
\label{fig:typerules}
\end{figure}

The next rule describes the multi-set representation for arrays. First, since arrays and collections are often used interchangably, we also represent arrays using the attribute {\tt collection}. Second, if $\tau$ is an array with element type $\tau'$, observe that  $\psi(\tau)$ also includes  the multiset representation of $\tau'$.  For example, $\psi(${\tt int[][]}$)$
is given by: $$\{ \tt int, numeric, Array, collection, Array, collection\}$$



The next two rules labeled {\it Wildcard}  and {\it TypeParam} allow us to handle generic types. Specifically, we define $\psi(\tau) = \emptyset$ for both the wild card type {\tt ?} as well as a type parameter, such as {\tt E}. For parametrized types $\tau$ (e.g., {\tt List<int>}), the \emph{Parametrized} rule defines $\psi(\tau)$ to be   the union of the multiset representation of the declaring raw type as well as its type arguments.

%

\begin{example}
The multiset representation of a generic list, ${\tt List \langle E \rangle}$, is  $ \psi({\tt List}) \cup \psi({\tt E}) =$ ${\tt \{List, collection\}}$. The multiset representation of a parameterized vector, such as ${\tt Vector \langle Vector \langle Integer \rangle \rangle}$, is $\psi({\tt Vector \langle Vector \langle Integer \rangle \rangle})$ $=$ $\psi({\tt Vector}) \cup \psi({\tt Vector}) \cup \psi({\tt int}) =$ {\tt \{Vector, collection, Vector, collection,} {\tt int,} {\tt nu\-meric\}}. 
\end{example}

For reference types (i.e., user-defined classes), we perform one-level unrolling for the fields using the rules  {\it RefType} and {\it FieldList I - III}. For example, given a type {\tt Point} with two  fields of type {\tt int}, we have $\psi({\tt Point)} = {\tt Point} \cup \psi(int) \cup \psi(int)$. Similarly, for a recursive type defined as:

{\small
\vspace{2mm}
\begin{myverb}  
class ListNode { int key; ListNode next; },
\end{myverb}
\vspace{2mm} 
}

\noindent we have $\psi({\tt ListNode}) = {\tt \{ListNode, int, numeric, ListNode\}}$.

\subsection{Computing Distances Between Types}

We now consider how to compute the \emph{distance} between a pair of types from their multiset representation. Given two types  $\tau$ and $\tau'$, we write $\delta(\tau, \tau')$ to denote the distance between $\tau$ and $\tau'$ and define it to be the normalized cardinality of  the {\it symmetric difference} between $\psi(\tau)$ and $\psi(\tau')$. More formally, $\delta(\tau, \tau')$ is defined as follows:

\begin{definition}[Type distance]  
\begin{equation*}
\delta(\tau, \tau') = \frac{|(\psi(\tau) - \psi(\tau')) \cup (\psi(\tau') - \psi(\tau))|}{|\psi(\tau) \cup \psi(\tau')|}
\end{equation*}
\end{definition}

Intuitively,  $\delta(\tau, \tau') \in [0,1]$  represents the fraction of elements in $\psi(\tau) \cup \psi(\tau')$ that are not shared between $\psi(\tau)$ and $\psi(\tau')$. Hence, $\delta(\tau, \tau')$ is the normalized cardinality of  the {\it symmetric difference} between $\psi(\tau)$ and $\psi(\tau')$.   Observe that, if $\tau$ and $\tau'$ are the same type, then  $\delta(\tau, \tau')$ is always zero. Also observe that $\delta(\tau, \tau')$ is always the same as $\delta(\tau', \tau)$.  Furthermore, if $\delta(\tau, \tau_1) < \delta(\tau, \tau_2)$, this indicates that $\tau_1$ is more similar to $\tau$ than $\tau_2$ is to $\tau$. 





\begin{example}
Consider the types $\tau_1$: {\tt ArrayList<Integer>}, $\tau_2$: {\tt LinkedList<Double>}, $\tau_3$: {\tt HashSet<Double>}
and their corresponding multiset representations:

\begin{center}
\begin{tabular}{l}
$\psi(\tau_1)$: \{{\tt int, numeric, List, collection}\} \\
$\psi(\tau_2)$: \{{\tt double, numeric, List, collection}\} \\
$\psi(\tau_3)$: \{{\tt double, numeric, Set, collection}\} \\
\end{tabular}
\end{center}

We have $\delta(\tau_1, \tau_2) = 0.25$ and $\delta(\tau_1, \tau_3) = 0.5$. Hence, $\tau_2$ is more similar to $\tau_1$ compared to $\tau_3$.


\end{example}
In the rest of this paper, we will slightly abuse this notation and write $\delta(x, y)$ to denote the distance between the \emph{types} of variables $x$ and $y$.

\newcommand{\mapping}{\mathcal{M}}
\newcommand{\indicator}{\mathbb{I}}

\section{Interface alignment}
\label{sec:mapping}

We now define the \emph{interface alignment} problem and show how to find an optimal alignment using integer linear programming. 

\subsection{Problem Definition}

Consider an adapter method with name $\adapter$ and type signature $\tau_1 \times \ldots \times \tau_n \rightarrow \tau$. 
Also, suppose that we have the source code of an adaptee method with name $\adaptee$ and signature $\tau'_1 \times \ldots \times \tau'_m \rightarrow \tau'$. In order to synthesize an implementation of $\adapter$  using $\adaptee$ as an adaptee, we must first identify a candidate mapping between the parameters of $\adapter$ and those of $\adaptee$. For example, the first argument of $\adaptee$ may correspond to the second parameter of $\adapter$. 
Furthermore, $\adapter$ and $\adaptee$ need not necessarily have the same number of parameters. For example, if $\adapter$ has signature  ${\tt Point} \rightarrow {\tt void}$ and  $\adaptee$ has signature ${\tt int} \times {\tt int} \rightarrow {\tt void}$, we may need to use both parameters of $\adaptee$ to construct the first argument of $\adapter$. Hence, a precursor to synthesizing adapter code is to identify a suitable mapping between the parameters of $\adapter$ and $\adaptee$. We refer to this problem as \emph{interface alignment}.

\begin{definition}[Interface alignment]
Consider  adapter  $\adapter$ and adaptee $\adaptee$ with the following signatures:
\footnotesize
\begin{equation*}
\begin{aligned}
\mathcal R &: r_1 : \tau_1 \times r_2 : \tau_2 \times \ldots \times r_n : \tau_n \to r_0 : \tau \\
\mathcal E &: e_1 : \tau_1' \times e_2 : \tau_2' \times \ldots \times e_m : \tau_m' \to e_0 : \tau'
\end{aligned}
\end{equation*}
\normalsize
where $x : \tau$ indicates  $x$ has type $\tau$. Let $P(\mathcal R)$ denote the set $\{r_1, r_2, \ldots, r_n, r_0 \}$ if the return value of $R$ is not {\tt void} and $\{r_1, r_2, \ldots, r_n \}$ otherwise.
The interface alignment problem is to find a  mapping $\mathcal M$ from $P(\mathcal E)$  to $ P(\mathcal R)$ such that (i) $\mapping$ is surjective, and (ii) $\mapping$ is \emph{not} many-to-many.
\end{definition}

Intuitively, we require $\mathcal{M}$ to be \emph{surjective} because all parameters of $\adapter$ should be used in $\adapter$'s implementation. Furthermore, we do not require $\mathcal{M}$ to be injective because different arguments of the adaptee can be mapped to the same argument in the adapter. For instance, for the motivating example from Section~\ref{sec:example}, both {\tt x1} and {\tt y1} from the adaptee are mapped to argument {\tt pt} of the adapter.

It is important to emphasize that a solution  $\mathcal{M}$ to the interface alignment problem  does not need to be a function. In particular, we allow many-to-one mappings. To see why this is necessary,  consider an adaptee that has a single argument {\tt pt} of type {\tt Point} and an adapter that takes two arguments {\tt x, y} of type {\tt int}. In this case, since we can construct a {\tt Point} from two {\tt int}'s, a suitable alignment maps {\tt pt} to both {\tt x} and {\tt y}, meaning that we construct {\tt pt} from {\tt x} and {\tt y}.  Finally, since $\mathcal{M}$ does not have to be a function, some parameters of the adaptee may also be unassigned. For instance, for the motivating example from Section~\ref{sec:example}, parameters {\tt x0} and {\tt y0} of {\tt bresenham} are not assigned to any parameters of {\tt drawLine}. If a parameter {\tt x} of the adaptee is unassigned in $\mathcal{M}$, we synthesize a constant value of {\tt x} using a pre-defined set of constants for each type  (e.g., \{0, 1\} for {\tt int} and $\{ \emph{true}, \emph{false} \}$ for {\tt boolean}).

Observe that a given solution $\mathcal{M}$ to the interface alignment problem  describes a particular strategy for generating wrapper code. In particular, if $\mathcal{M}(x:\tau) = \{ y_1:\tau_1, \ldots, y_n:\tau_n \} $, this means that we should construct parameter $x$ of the adaptee using parameters $y_1, \ldots, y_n$ of the adapter. Our approach uses an existing type-directed synthesis tool~\cite{sypet} to generate code snippets that produce an object of type $\tau$ using objects of type $\tau_1, \ldots, \tau_n$.

\subsection{Optimal Interface Alignment}

Given an adapter $\adapter$ and a candidate adaptee $\adaptee$, there are, in general, \emph{many} possible solutions to the interface alignment problem. One of the key insights underlying our approach is to use type distances to find a \emph{good} alignment. Given a  mapping $\mathcal{M}$, we define $\emph{Cost}(\mathcal{M})$ as follows:

\begin{definition}[Cost of mapping]
\label{def:mapping}
Let $\indicator(x)$ be 1 if $|M(x)| \neq 1$ and 0 otherwise. Then, the \emph{cost} of $\mathcal{M}$ is given by:~\footnote{If $\mathcal{M}(x)$ is not a singleton, we treat $\mathcal{M}(x)$ as a list and compute $\delta(x, \mathcal{M}(x))$ using the \emph{FieldList} rules from Figure~\ref{fig:typerules}.}
\[
\begin{array}{cc}
\emph{Cost}(\mathcal{M}) =  & \left ( 
\begin{array}{l} \sum_{x \in \emph{dom}(\mathcal{M})} \indicator(x) \cdot \delta(x,  \mathcal{M}(x)) \  + \\
\sum_{y \in \emph{dom}(\mathcal{M}^{-1})} (1-\indicator(y)) \cdot \delta(y,  \mathcal{M}^{-1}(y))
\end{array}
\right )
\end{array}
\]
\end{definition}

To understand the rationale behind this definition, consider the case where $\mapping$ is a many-to-one mapping (e.g., $x:int \mapsto pt:Point, y:int \mapsto pt:Point$). In this case, an argument  (e.g., \emph{pt}) of $\adapter$ corresponds to multiple arguments (e.g., $x, y$) of the adaptee $\adaptee$. In our example, this means that we must ``convert" a point object to a pair of integers; hence, the cost of this mapping should be $\delta(\emph{Point}, \{ \emph{int, int} \})$ rather than $2*\delta(\emph{int}, \emph{Point})$.  Hence, intuitively, the cost of the mapping depends on whether $M$ is one-to-many or many-to-one. To capture this intuition, we weight each term $\delta(x, \mapping(x))$ with $\indicator(x)$, meaning that we should only pay the cost $\delta(x, \mapping(x))$ if $\mapping$ is one-to-many. Otherwise, we pay the cost $\delta(y, \mapping^{-1}(y))$ using the reverse mapping.

\begin{example}
Consider the following signatures:
\[
\begin{array}{ll}
\adaptee: & x:{\tt int} \times y:{\tt int} \rightarrow {\tt void} \\
\adapter: & p: {\tt Point} \rightarrow {\tt void}
\end{array}
\]
as well as the following mappings:
\[
\begin{array}{ll}
\mapping_1: & [x \mapsto \{ p\}, y \mapsto \{p \} ] \\
\mapping_2: & [x \mapsto \{ p\}, y \mapsto \emptyset ] 
\end{array}
\]
Here, we have $\emph{Cost}(\mapping_1) = \delta(p, \{x,y\})$. On the other hand, $\emph{Cost}(\mapping_2) = \delta(y, \emptyset) + \delta(p, x)$.
\end{example}

Based on this notion of cost, we can now define what it means for an alignment to be optimal:

\begin{definition}[Optimal alignment]
$\mathcal{M}^*$ is an \emph{optimal} solution to the interface alignment problem if there does not exist another solution $\mathcal{M}$ such that $\emph{Cost}(\mathcal{M}) < \emph{Cost}(\mathcal{M^*})$.
\end{definition}

In other words, an optimal solution  minimizes the cost of the mapping according to the type distance matrix.  In the next section, we show how to find an optimal solution using 0-1 Integer Linear Programming.

 
\subsection{ILP Formulation}

A 0-1 Integer Linear Programming (ILP) 
consists of a set of linear constraints $\mathcal C$ over boolean variables  and an objective function $c$. The goal is to find an assignment  such that all constraints are satisfied and the value of the objective function $c$ is optimized. 

\begin{definition}[0-1 Integer Linear Programming]
The 0-1 ILP problem is defined as follows:

\begin{equation*}
\begin{aligned}
min~c : \sum_j c_j x_j \ \ s.t. \ \ ~\mathcal C : \bigwedge_i \sum_j a_{i,j} x_j~\Delta~b_i, 
\end{aligned}
\end{equation*}
with $\Delta := \{\leq, =, \geq\}$, and $x_j \in \{0,1\}$.
\end{definition}

We formulate the problem of finding an optimal interface alignment using 0-1 ILP. Specifically, constraints $\mathcal{C}$ encode that $\mathcal{M}$ is a surjective mapping from $P(\adaptee)$ to $P(\adapter)$ and that it is not many-to-many. The objective function expresses that we want to minimize the cost of this mapping. In what follows, we describe our encoding in more detail.


\vspace{0.1in} \noindent
{\bf\emph{Variables.}} The variables in our 0-1 ILP formulation correspond to all possible mappings from $P(\adaptee)$ to $P(\adapter)$.
In particular, a boolean variable $x_{e \rightarrow \vec{r}}$ indicates a one-to-many mapping from  parameter $e \in P(\adaptee)$ to a set of parameters $\vec{r} \subseteq P(\adapter)$.~\footnote{We consider a one-to-one mapping a particular case of the one-to-many mapping where $\vec{r}$ only contains one parameter.} Similarly, boolean variables $x_{\vec{e} \rightarrow r}$ indicate a many-to-one mapping from parameters $\vec{e} \subseteq P(\adaptee)$ to parameter $r \in P(\adapter)$.
Note that only parameters of reference type are allowed to have many-to-one or one-to-many mappings: If two parameters $r_1, r_2 \in P(\adapter)$ are mapped to a parameter $e_1 \in P(\adaptee)$ of reference type, this means that $e_1$ is constructed using $r_1$ and $r_2$. Similarly, if two parameters $e_1, e_2 \in P(\adaptee)$ are mapped to a single parameter $r_1 \in P(\adapter)$ of reference type, this means that $e_1, e_2$ are deconstructed using $r_1$ (e.g., via getter methods). Hence, while many-to-one and one-to-many mappings make sense for reference types, it does not have a sensible interpretation for primitive types and collections.

We use $\mathcal{V}$ to denote all variables used in the encoding. Given a parameter $r \in P(\adapter)$, $\mathcal{V}_r$ denotes the set of variables where $r$ occurs. Similarly, given a parameter $e \in P(\adaptee)$, $\mathcal{V}_e$ denotes the set of variables where $e$ occurs.



\begin{example}
\label{ex:variables}
Assume we have an adapter method $\adapter$ with signature $(r_1: {\tt Point} \times r_2: {\tt long}) \to (r_0: {\tt void})$ and an adaptee method $\adaptee$ with signature $(e_1: {\tt int} \times e_2: {\tt int} \times e_3: {\tt long}) \to (e_0: {\tt void})$. Then we have $P(\adaptee) = \{e_1, e_2, e_3\}$ and $P(\adapter) = \{r_1, r_2\}$. 
Thus, the set of variables are:

\begin{equation*}
\begin{aligned}
\mathcal V = \ & \{x_{e_1 \rightarrow r_1}, x_{e_1 \rightarrow r_2}, x_{e_2 \rightarrow r_1}, x_{e_2 \rightarrow r_2}, x_{e_3 \rightarrow r_1}, x_{e_3 \rightarrow r_2} \\
& x_{e_1,e_2 \rightarrow r_1}, x_{e_1,e_3 \rightarrow r_1}, x_{e_2,e_3 \rightarrow r_1}, x_{e_1,e_2,e_3 \rightarrow r_1}\}.\\
\end{aligned}
\end{equation*}

Given the parameter $e_1$, $\mathcal{V}_{e_1}$ denotes the set of variables where $e_1$ occcurs:
$$\mathcal{V}_{e_1} = \{x_{e_1 \rightarrow r_1}, x_{e_1 \rightarrow r_2}, x_{e_1,e_2 \rightarrow r_1}, x_{e_1,e_3 \rightarrow r_1}, x_{e_1,e_2,e_3 \rightarrow r_1}\}.$$


A mapping is defined by the variables that are assigned to 1. For instance, consider an assignment $\sigma$ that assigns $x_{e_1,e_2 \rightarrow r_1}$ and $x_{e_3 \rightarrow r_2}$ to 1, and all other variables to 0.  Then, $\sigma$ corresponds to the following mapping: $(e_1, e_2) \to (r_1)$, $(e_3) \to (r_2)$.
\end{example}

Observe that the number of variables used in the encoding grows quadratically for non-reference types and exponentially for reference types. However, since the number of parameters is usually small, our encoding  introduces a manageable number of variables in practice.

\vspace{0.1in} \noindent
{\bf\emph{Constraints.}}  While the variables describe all possible mappings between parameters of an adaptee  $\adaptee$ and a desired method $\adapter$, not all mappings can occur simultaneously. In particular, we must enforce that any satisfying assignment to $\mathcal{C}$ corresponds to a surjective mapping from $P(\adaptee)$ to $P(\adapter)$. Furthermore,  types also impose \emph{hard constraints} that limit which variables in $P(\adaptee)$ can be mapped to which ones in $P(\adapter)$. 
We enforce these hard constraints by generating a system of linear constraints $\mathcal C$ as follows:


\begin{enumerate}
\item First, we divide all Java types into 3 categories, namely \emph{primitive}, \emph{collection}, and \emph{reference}, and only allow  type conversions marked as ``Y'' in the following table:

    \begin{center}
    \begin{tabular}{| c | c | c | c |}
    \hline
      & primitive & collection & reference \\
    \hline
    primitive & Y & N & Y \\
    \hline
    collection & N & Y & N \\
    \hline
    reference & Y & N & Y \\
    \hline
    \end{tabular}
    \end{center}

    If two parameters $r \in P(\adapter)$ and $e \in P(\adaptee)$ are not compatible due to their types, the boolean variables where these parameters occur are always set to 0:
    
    {\small
    \begin{equation*}
    \begin{aligned}
    &x_{e \rightarrow \vec{r}} = 0 \ \  {\rm if} \ \  x_{e \rightarrow \vec{r}} \in (\mathcal{V}_e \cap \mathcal{V}_r), \text{({\tt type}(r), {\tt type}(e)) is ``N''}\\
    &x_{\vec{e} \rightarrow r} = 0 \ \  {\rm if} \ \  x_{\vec{e} \rightarrow r} \in (\mathcal{V}_e \cap \mathcal{V}_r), \text{({\tt type}(r), {\tt type}(e)) is ``N''}
    \end{aligned}
    \end{equation*}
    }
We note that the above constraints are somewhat connected with the power of the synthesis tool used for generating wrapper code. In particular, we disallow mappings between references and collections since \system's underlying synthesis engine does not support converting between references and collections.

\item For each parameter $r \in P(\adapter)$, we impose that there is {\it exactly one} mapping that contains $r$:
\begin{equation*}
\forall {r \in P(\adapter)}. \  \sum_{x_{e \rightarrow \vec{r}} \in \mathcal{V}_r} x_{e \rightarrow \vec{r}} \  + \sum_{x_{\vec{e} \rightarrow r} \in \mathcal{V}_r} x_{\vec{e} \rightarrow r} = 1
\end{equation*}

Effectively, these constraints enforce that any solution of $\mathcal{C}$ corresponds to a surjective mapping.

\item For each parameter $e \in P(\adaptee)$, we impose that there is {\it at most one} mapping that contains $e$:
\begin{equation*}
\forall {e \in P(\adaptee)}. \  \sum_{x_{e \rightarrow \vec{r}} \in \mathcal{V}_e} x_{e \rightarrow \vec{r}} \ + \sum_{x_{\vec{e} \rightarrow r} \in \mathcal{V}_e} x_{\vec{e} \rightarrow r} \leq 1
\end{equation*}

These constraints enforce that a parameter $e \in P(\adaptee)$ can only be used in at most one mapping. In particular, it is not necessary that every $e \in  P(\adaptee)$ be mapped to some $r \in P(\adapter)$.

\end{enumerate}


\begin{example}
\label{ex:constraints}
Consider the same methods $\adapter$ and $\adaptee$ from Example~\ref{ex:variables}. Constraints $\mathcal C$ for this example are given by:

{\small
\vspace{-0.1in}
\begin{equation*}
\resizebox{.99\hsize}{!}{
$\begin{aligned}
p_1 :&~ x_{e_1 \rightarrow r_1} + x_{e_2 \rightarrow r_1} + x_{e_3 \rightarrow r_1} + x_{e_1,e_2 \rightarrow r_1} + \\
     &~ x_{e_1,e_3 \rightarrow r_1} + x_{e_2,e_3 \rightarrow r_1} + x_{e_1,e_2,e_3 \rightarrow r_1} = 1\\
p_2 :&~ x_{e_1 \rightarrow r_2} + x_{e_2 \rightarrow r_2} + x_{e_3 \rightarrow r_2} = 1\\
p_3 :&~ x_{e_1 \rightarrow r_1} + x_{e_1 \rightarrow r_2} + x_{e_1,e_2 \rightarrow r_1} + x_{e_1,e_3 \rightarrow r_1} + x_{e_1,e_2,e_3 \rightarrow r_1} \leq 1\\
p_4 :&~ x_{e_2 \rightarrow r_1} + x_{e_2 \rightarrow r_2} + x_{e_1,e_2 \rightarrow r_1} + x_{e_2,e_3 \rightarrow r_1} + x_{e_1,e_2,e_3 \rightarrow r_1} \leq 1\\
p_5 :&~ x_{e_3 \rightarrow r_1} + x_{e_3 \rightarrow r_2} + x_{e_1,e_3 \rightarrow r_1} + x_{e_2,e_3 \rightarrow r_1} + x_{e_1,e_2,e_3 \rightarrow r_1} \leq 1\\
\end{aligned}$
}
\end{equation*}
}

Here, $p_1$ and $p_2$ enforce that each parameter $r \in P(\adapter)$ appears in exactly one mapping. Constraints $p_3$, $p_4$, $p_5$ guarantee that each parameter $e \in P(\adaptee)$ appears in at most one mapping. These constraints 
enforce that any solution to $\mathcal{C}$ corresponds to a surjective mapping.
\end{example}


\vspace{0.1in} \noindent
{\bf\emph{Objective function.}} The goal of the objective function $c$ 
is to find an \emph{optimal} alignment with the lowest cost. Specifically, we define the objective function $c$ 
as follows:

\begin{equation*}
  \sum_{e \in P(\adaptee)} \sum_{x_{e \rightarrow \vec{r}} \in \mathcal{V}_e} \delta(e, \vec{r}) \cdot x_{e \rightarrow \vec{r}} \ ~+ \sum_{r \in P(\adapter)}  \sum_{x_{\vec{e} \rightarrow r} \in \mathcal{V}_r} \delta(r, \vec{e}) \cdot x_{\vec{e} \rightarrow r}
\end{equation*}

Each mapping has an associated cost using type distances from Section~\ref{sec:distance}. Observe that this objective function directly encodes the cost metric from Definition~\ref{def:mapping}. 

\begin{example}
Consider the same methods $\adapter$ and $\adaptee$ from Example~\ref{ex:variables} and the constraints from Example~\ref{ex:constraints}. Here, we have a total of 9 valid solutions to the constraint system. Each of these mappings, along with their cost, is listed below:

\begin{center}
\resizebox{.85\hsize}{!}{
\begin{tabular}{c | l | l}
\hline
id & mapping & cost \\
\hline
$m_1$ & $(e_1, e_2) \to (r_1), (e_3) \to (r_2)$ & $\sfrac{1}{9} + 0 \simeq 0.11$\\
$m_2$ & $(e_1) \to (r_1), (e_3) \to (r_2)$ & $\sfrac{3}{7} + 0 \simeq 0.43$ \\
$m_3$ & $(e_2) \to (r_1), (e_3) \to (r_2)$ & $\sfrac{3}{7} + 0 \simeq 0.43$ \\
$m_4$ & $(e_1, e_3) \to (r_1), (e_2) \to (r_2)$ & $\sfrac{3}{9} + \sfrac{2}{4} \simeq 0.83$ \\
$m_5$ & $(e_2, e_3) \to (r_1), (e_1) \to (r_2)$ & $\sfrac{3}{9} + \sfrac{2}{4} \simeq 0.83$ \\
$m_6$ & $(e_1) \to (r_1), (e_2) \to (r_2)$ & $\sfrac{3}{7} + \sfrac{2}{4} \simeq 0.93$ \\
$m_7$ & $(e_2) \to (r_1), (e_1) \to (r_2)$ & $\sfrac{3}{7} + \sfrac{2}{4} \simeq 0.93$ \\
$m_8$ & $(e_3) \to (r_1), (e_1) \to (r_2)$ & $\sfrac{5}{7} + \sfrac{2}{4} \simeq 1.21$ \\
$m_9$ & $(e_3) \to (r_1), (e_2) \to (r_2)$ & $\sfrac{5}{7} + \sfrac{2}{4} \simeq 1.21$ \\
\hline
\end{tabular}
}
\end{center}

Since our objective function is to minimize the cost, the first result we  get is $m_1$, which is exactly what we want.
\end{example}

\begin{table*}[!t]
\begin{center}
\scalebox{0.9}{
\begin{tabular}{lrr}
\hline
& \system & \ssix\\
\hline
Success rate: & 100.0\% & 37.5\%\\
\hline
\end{tabular}
}
\end{center}
\vspace{-2mm}
\centering
\scalebox{0.9}{
\begin{tabular}{llrr}
\hline
Id & Benchmark & \system & \ssix\\
\hline
1 & SimpleTokenizer & \cmark & \cmark\\
2 & QuoteTokenizer & \cmark & \cmark\\
3 & CheckRobots & \cmark & \cmark\\
4 & LogBase & \cmark & \cmark\\
5 & RomanNumeral & \cmark & \cmark\\
6 & RomanToInt & \cmark & \cmark\\
7 & PrimeNumber & \cmark & \cmark\\
8 & PerfectNumber & \cmark & \cmark\\
9 & DayOfWeek & \cmark & \cmark\\
10 & EasterDate & \cmark & \cmark\\
11 & MatrixMultiplication & \cmark & \xmark \\
12 & LcsInteger& \cmark & \xmark \\
13 & RemoveDuplicates& \cmark & \cmark \\
14 & TransposeMatrix& \cmark & \xmark \\
15 & InvertMatrix& \cmark & \xmark \\
16 & MatrixPower& \cmark & \xmark \\
17 & DotProduct& \cmark & \xmark \\
18 & MatrixDeterminant& \cmark & \cmark \\
19 & CountingSort& \cmark & \xmark \\
20 & MatrixAddition& \cmark & \xmark \\
\hline
\end{tabular}
\hspace*{2cm}
\begin{tabular}{llrr}
\hline
Id & Benchmark & \system & \ssix\\
\hline
21 & FloodFill& \cmark & \xmark \\
22 & FindMedian& \cmark & \xmark \\
23 & ListAverage& \cmark & \xmark \\
24 & BresenhamLine& \cmark & \xmark \\
25 & BresenhamCircle& \cmark & \xmark \\
26 & Distance& \cmark & \xmark \\
27 & Slope& \cmark & \xmark \\
28 & PrimeSieve& \cmark & \cmark \\
29 & Anagram& \cmark & \cmark \\
30 & Palindrome& \cmark & \cmark \\
31 & PartitionList& \cmark & \xmark \\
32 & RotateList& \cmark & \xmark \\
33 & ListInsertion& \cmark & \xmark \\
34 & PalindromeList& \cmark & \xmark \\
35 & SwapNodesPairs& \cmark & \xmark \\
36 & InvertBinaryTree& \cmark & \xmark \\
37 & MinDepthBinaryTree& \cmark & \xmark \\
38 & BinaryTreePostorder& \cmark & \xmark \\
39 & BinaryTreeInorder& \cmark & \xmark \\
40 & SumRootLeafNumbers& \cmark & \xmark \\
\hline
\end{tabular}
}
\vspace{-1mm}
\caption{Comparison between \system and \ssix}
\label{tbl:benchmarks}
\end{table*}

\section{Design and Implementation}
In this section, we describe the design and implementation of the \system tool for automated code reuse.  \system is publicly available as an Eclipse plug-in at the Eclipse marketplace 
and consists of approximately 12,000 
lines of Java code. Our implementation uses the Sat4J~\cite{sat4j} tool for solving 0-1 ILP problems.
%
%

\system can be integrated with any code search engine that yields results at the granularity of methods. In our current implementation, we integrated \system with the Pliny code search engine~\cite{pliny} and use a database of over 12 million  Java methods collected from open source repositories, such as Github~\cite{github} and Bitbucket~\cite{bitbucket}.



\vspace{0.1in} \noindent
{\bf \emph{Re-ranking.}} Given the initial results provided by the Pliny search engine, \system re-ranks search results using the type similarity matrix described in Section~\ref{sec:distance}. Specifically, for each method $M$ in the search result, \system computes a \emph{type similarity score} between the signature of $M$ and that of the desired method $M^*$. The type similarity score is the multiplicative inverse of the cost for the optimal alignment between $M$ and $M^*$. Hence, our re-ranking procedure sorts the original search results using the techniques described in Section~\ref{sec:mapping} for finding an optimal alignment. As we demonstrate in Section~\ref{sec:eval}, our re-ranking algorithm based on type similarity significantly improves search results.

\vspace{0.1in} \noindent
{\bf \emph{Synthesis of wrapper code.}} As mentioned earlier, a solution to the interface alignment problem can be immediately translated into wrapper code using existing synthesis tools~\cite{sypet,jungloid,insynth,codehint}. Our current implementation utilizes the {\sc SyPet} tool for type-directed component-based synthesis~\cite{sypet}. Specifically, let $\mapping$ be a solution to the interface alignment problem, and let $\mapping(x:\tau)$ be $\{y_1:\tau_1, \ldots, y_n:\tau_n \}$. In this case, to synthesize the actual value for $x$, we ask {\sc SyPet} to generate a procedure that takes inputs of type $\tau_1, \ldots, \tau_n$ and produces an output of type $\tau$.  Since there can be many ways to construct an object of type $\tau$ from $y_1, \ldots, y_n$, our implementation tries the top 100 results returned by {\sc SyPet} before moving on to a different solution to the interface alignment problem (this limit was never reached during our experiments).

Since most type-directed synthesis tools, including {\sc SyPet}, cannot synthesize loops,  we have developed a library of templates for performing conversions between different types of collections. For example, to generate a collection $x$ of type {\tt Vector<Foo>} from another collection $y$ of type {\tt Bar[]}, we first retrieve the built-in template for converting an array to a vector and then invoke the procedure synthesized by {\sc SyPet} on each element. 

As mentioned earlier, \system can also synthesize constants in cases where an adaptee parameter is mapped to the empty set. In our current implementation, each type has a set of ``default" values, including {\tt NULL} for references, $\{\emph{true,false}\}$ for booleans, and $\{0,1\}$ for integers. If an adaptee parameter $x$ of type $\tau$ is mapped to $\emptyset$, \system tries all default values associated with type $\tau$.

\vspace{0.1in} \noindent
{\bf \emph{Dependency resolution.}} After synthesizing wrapper code for a candidate alignment, \system resolves all dependencies between classes based on the package hierarchy. Specifically, \system starts from the class $C$ containing the candidate method and transitively identifies all classes $C_1, \ldots, C_n$ that $C$ depends on. Since this procedure can take a long time for large dependencies, we restrict the maximum number of classes to 10.  If the dependencies of a candidate cannot be fully resolved within this limit, \system backtracks and tries the next best candidate.

\section{Evaluation}
\label{sec:eval}

To evaluate the usefulness of \system, we perform three sets of experiments. First, we compare \system against \ssix~\cite{s6}, a state-of-the-art code reuse tool that is publicly available and well-maintained. Second, we use \system to re-rank the search results provided by the Pliny~\cite{pliny} and Grepcode~\cite{grepcode} search engines and compare the results before and after re-ranking. 
Finally, we perform a user study and evaluate how long participants take to complete various algorithmic tasks with and without \system. 
All experiments are conducted on a Lenovo laptop with an 
Intel i7-5600U CPU and 8G of memory running Ubuntu 14.04. 

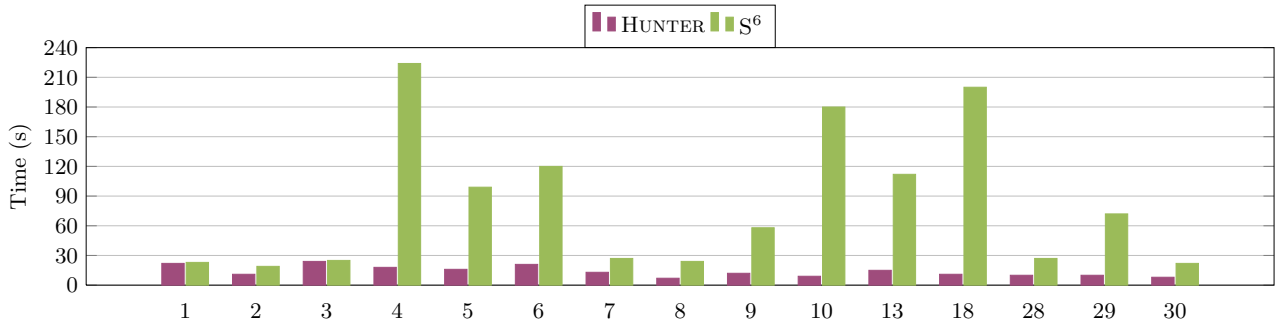
\begin{figure*}[!t]
\centering
\scalebox{0.94}{
\hspace*{-3mm}
\begin{tikzpicture}
    \begin{axis}[
        y=0.014cm,
        x=1cm,
        major x tick style = transparent,
        ybar=2*\pgflinewidth,
        bar width=9pt,
        ymajorgrids = true,
        ylabel = {Time (s)},
        ylabel style={yshift=-3mm},
        symbolic x coords={1,2,3,4,5,6,7,8,9,10,13,18,28,29,30},
        xtick = data,
        ytick={0,30,60,90,120,150,180,210,240},
        scaled y ticks = false,
        ymin=0,
        ymax=240,
        legend cell align=left,
        legend entries={\system, \ssix, No solution},
        legend style={
                at={(0.5,1.181)},
                legend columns=-1,
                anchor=north,
        }
    ]
        \addplot[style={ppurple,fill=ppurple,mark=none}]
         coordinates{(1,22)(2,11)(3,24)(4,18)(5,16)(6,21)(7,13)(8,7)(9,12)(10,9)(13,15)(18,11)(28,10)(29,10)(30,8)};

        \addplot[style={ggreen,fill=ggreen,mark=none}]
         coordinates{(1,23)(2,19)(3,25)(4,224)(5,99)(6,120)(7,27)(8,24)(9,58)(10,180)(13,112)(18,200)(28,27)(29,72)(30,22)};

\end{axis}
\end{tikzpicture}
}
\vspace{-0.12in}
\caption{Comparison of running times between \system and \ssix}
\label{fig:s6}
\end{figure*}

\subsection{Comparison with S$^6$}
\label{sec:comparison-s6}

\ssix~\cite{s6} is a state-of-the-art code reuse tool with a web interface that 
takes the same set of inputs as \system (i.e., method signature, natural language description, and test cases). There are two key differences between \ssix and \system. First,  \ssix directly modifies the search result instead of synthesizing wrapper code. 
Second, \system uses integer linear programming to measure similarity between type signatures and uses this information to re-rank search results and facilitate integration with synthesis tools.

We compare \system against \ssix on a variety of benchmarks collected from three different sources:  (i) examples used to evaluate \ssix~\cite{s6}, (ii) challenge problems taken from the Mining and Understanding Software Enclaves (MUSE) project~\cite{muse} 
and (iii) benchmarks from Leetcode~\cite{leetcode}, which is an online resource containing programming problems. 

Table~\ref{tbl:benchmarks} describes the 40 benchmarks used in our evaluation in more detail. In this table, a \cmark\ indicates that the tool was able to successfully synthesize the desired code, and an \xmark\ indicates that synthesis failed.  As summarized in Table~\ref{tbl:benchmarks}, \system is able to solve \emph{all} benchmarks with an average running time of 14 seconds. In contrast, \ssix can only solve 37.5\% of the benchmarks. 
%
Since both tools use the same underlying code corpus, these results indicate that \system is more successful at reusing existing code compared to \ssix.

To further investigate running time, Figure~\ref{fig:s6} compares the performance of \system against that of \ssix on the 15 benchmarks that could be successfully solved by both systems. The numbers on the x-axis refer to the numeric identifier of each benchmark, as presented in Table~\ref{tbl:benchmarks}. The y-axis shows the corresponding time that each tool took to solve those benchmarks. As we can see from Figure~\ref{fig:s6}, \system is also much faster compared to \ssix. On average, \system takes 15 seconds to solve these benchmarks, while \ssix takes an average of 83 seconds.

\subsection{Effect of Re-ranking on Search Results}
In this paper, we have argued that type similarity metrics can be used to improve the quality of results returned by code search engines. To substantiate this claim, we perform an experiment to  compare the quality of search results before and after using our signature-based re-ranking algorithm. Recall that our re-ranking technique  sorts search results based on the cost of the optimal solution to the interface alignment problem. If two methods $m, m'$ have the same cost according to our signature similarity metric, our re-ranking algorithm preserves the relative ordering between $m, m'$ based on the original search results.

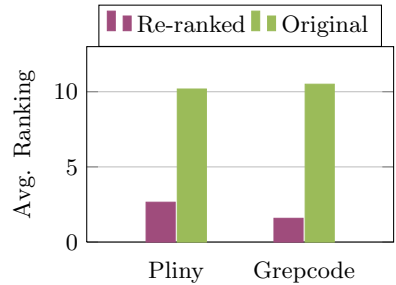
\begin{figure}[!th]
\centering
\begin{tikzpicture}
    \begin{axis}[
        y=0.2cm,
        x=1.7cm,
        major x tick style = transparent,
        ybar=2*\pgflinewidth,
        bar width=11pt,
        ymajorgrids = true,
        ylabel = {Avg. Ranking},
        ylabel style={yshift=-4mm},
        symbolic x coords={Pliny,Grepcode},
        xtick = data,
        scaled y ticks = false,
        ytick={0,5,10},
        ymin=0,
        ymax=13,
        enlarge x limits=0.7,
        legend cell align=left,
        legend entries={Re-ranked, Original},
        legend style={
                at={(0.5,1.215)},
                legend columns=-1,
                anchor=north,
        }
    ]


    \addplot[style={ppurple,fill=ppurple,mark=none}]
         coordinates{(Grepcode,1.58) (Pliny,2.65)};


    \addplot[style={ggreen,fill=ggreen,mark=none}]
    coordinates{(Grepcode,10.5) (Pliny,10.18)};

    \end{axis}
\end{tikzpicture}
\vspace{-0.15in}
\caption{Average ranking of solution}
\label{fig:reranking}
\end{figure}

To evaluate the effectiveness of our re-ranking algorithm,   we compare the ranking of the desired solution before and after re-ranking~\footnote{If there are multiple solutions that match a given query, we consider the rank of the highest rank solution. Furthermore, we manually inspected the search results to decide whether a given method implements the desired functionality.}. To show that this improvement is orthogonal to the code search engine, we perform the same experiment using two different code search engines, namely Pliny~\cite{pliny} and Grepcode~\cite{grepcode}. Since Grepcode does not use the same database as Pliny, we were only able to find the correct solution for 24 out of the 40 benchmarks from Table~\ref{tbl:benchmarks}. Hence, for Grepcode, we only report the ranking of the solution for the 24 benchmarks. For Pliny,  we report the rank of the solution before and after re-ranking for all 40 benchmarks.

Figure~\ref{fig:reranking} summarizes the results of this experiment. The green bars (on the right) show the rank of the solution in the original search results, and the purple bars (on the left) indicate the rank of the solution after re-ranking. Without using our re-ranking algorithm,  the solution has an average rank of 10 using Pliny and 11 using Grepcode.  On the other hand, after re-ranking, the solution has an average rank of 3 for Pliny and 2 for Grepcode. Hence, we believe these results indicate that our signature similarity metric can be used to improve the  results of any code search engine that yields results at the granularity of methods.

\subsection{User Study}
To evaluate the impact of \system on programmer productivity, we conducted a user study involving 16 graduate students, 2 professional programmers, and 3 undergraduate students. Among these 21 participants, 13 are experienced Java programmers, and the remaining 8 have limited exposure to Java. 


In our user study, we asked the participants to produce a Java implementation for each of the  following methods:

\begin{enumerate}
\item[$T_1$.] Double matrix multiplication:\\
{\small
\begin{myverb}
void multiply(Vector<Vector<Double>> first, 
              Vector<Vector<Double>> second,
              Vector<Vector<Double>> res)
\end{myverb}
}
\item[$T_2$.] Integer longest common subsequence:\\
\begin{myverb}
void lcs(Vector<Integer> first, 
         Vector<Integer> second, Vector<Integer> out)
\end{myverb}
\item[$T_3$.] Bresenham Line:\\
{\small
\begin{myverb}
void drawLine(int x, int y, Vector<MyPoint> p)
\end{myverb}
}
\end{enumerate}

For each problem, we provided a detailed description of the task to be performed, including a simple input-output example. Since our goal is to compare programmer productivity with and without \system, we instructed participants to complete each task using two different methods:

\begin{itemize}
\item {\bf \emph{Manual:}} In this scenario, participants were asked to complete the programming task manually without using \system. However, participants were explicitly told that they can use any existing code search tool and adapt the search results to their needs. In other words, the participants were aware that they do \emph{not} have to implement the algorithm from scratch.
\item  {\bf \emph{Using Hunter:}} In this scenario, participants were asked to use \system to automatically synthesize the desired code by providing a natural language query and implementing appropriate JUnit tests. Since participants did not have any prior familiarity with \system, we gave the participants a brief demo of the tool prior to the user study.
\end{itemize}

For both scenarios, we asked participants to stop working on a given task after 30 minutes. Hence, any task that the users could not complete within 30 minutes  was considered as a ``failure". For the manual implementation case, we did not require participants to write test cases for their code. Hence, there was no overlap between the tasks that the users needed to complete for the two different usage scenarios.

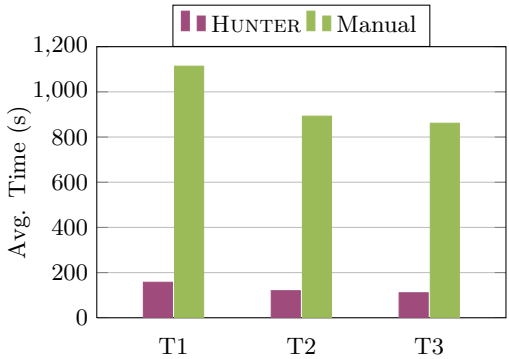
\begin{figure}[!t]
\centering
\begin{tikzpicture}
    \begin{axis}[
        y=0.003cm,
        x=1.7cm,
        major x tick style = transparent,
        ybar=2*\pgflinewidth,
        bar width=11pt,
        ymajorgrids = true,
        ylabel = {Avg. Time (s)},
        ylabel style={yshift=-2mm},
        symbolic x coords={T1,T2,T3},
        xtick = data,
        scaled y ticks = false,
        ytick={0,200,400,600,800,1000,1200},
        ymin=0,
        ymax=1200,
        enlarge x limits=0.3,
        legend cell align=left,
        legend entries={\system, Manual, Avg. \system, Avg. Manual},
        legend style={
                at={(0.5,1.155)},
                legend columns=-1,
                anchor=north,
        }
    ]


        \addplot[style={ppurple,fill=ppurple,mark=none}]
         coordinates{(T1,157.3) (T2,119.8) (T3,110.7)};

        \addplot[style={ggreen,fill=ggreen,mark=none}]
         coordinates{(T1,1113.6) (T2,892.4) (T3,861.15)};






    \end{axis}
\end{tikzpicture}
\vspace{-0.1in}
\caption{Comparison between \system and manual coding}
\label{fig:user}
\end{figure}
\vspace{-0.1in}

\vspace{0.1in} \noindent
{\bf \emph{Programmer productivity.}} Figure~\ref{fig:user} shows the average time for completing each of the three tasks with and without \system. Overall, participants took an average of 130 seconds using \system and an average of 948 seconds when writing the code manually. Furthermore, while participants were able to successfully complete 100\% of the tasks  using \system,  success rate was only 85\% without \system. In particular, participants were not able to manually complete 15\% of the tasks (4 $T_1$, 4 $T_2$, and 2 $T_3$) within the 30 minute time limit. 
In summary, these results demonstrate that \system allows programmers to be more productive. 

To further validate our claim that these results are statistically significant, we also performed  a two-tailed paired t-test~\cite{ttest} for each task, ignoring  samples that were not completed within the 30 minute time-limit. Since the t-test for each task returned p-values smaller than $0.0001$, this evaluation supports our claim that there is a significant difference in average task completion time with and without \system.


\vspace{0.1in} \noindent
{\bf \emph{Quality of solutions.}} 
To compare the \emph{quality} of the solutions implemented manually vs. using \system, we also manually inspected the solution and ran the programs on a large test suite.
While programs synthesized by \system pass all test cases, 19\% of the manually written programs
are actually buggy and fail at least one of our test cases. Among those buggy programs, 
ten of them 
contain off-by-one bugs, one program implementing task $T_1$
throws an exception for matrices that are not square, and one program 
implementing task $T_3$ produces the wrong result due to a typo in the copy-pasted code.
In summary, we believe these results show that programs synthesized by \system are less error-prone compared to their manually written versions.

\subsection{Threats to Validity}
In this section, we discuss possible threats to the validity of the experiments.

\vspace{1em}
\noindent {\bf\emph{Comparison with S$^6$.}}
One of the threats to validity in our evaluation is that we could only compare against a single test-driven code reuse tool, namely \ssix. In particular, we were not able to compare \system against 
 other test-driven code reuse tools, such as CodeGenie~\cite{codegenie} and CodeConjurer~\cite{codeconjurer}, since they are no longer maintained by the original developers.

Another threat to validity is the measurement of the running times when comparing \system and \ssix. Specifically, \ssix is multi-threaded, runs on a remote server, and it  terminates when it finds {\it all} methods that pass all test cases. \system is single-threaded, runs on a common laptop, and it terminates when it finds the {\it first} method that passes all test cases. We extended \system to terminate after finding all methods, and the average running time increases from 15 seconds to 25 seconds but it is still much faster compared to \ssix (83 seconds).

\vspace{1em}
\noindent {\bf\emph{Effect of re-ranking.}} 
Another threat to validity is that our re-ranking procedure requires search engines to produce results at the granularity of methods.  However, most search engines, such as Git\-hub~\cite{github}, OpenHub~\cite{openhub}, and CodeExchange~\cite{codeexchange}, provide solutions either at the granularity of files or require an exact match between method names. We chose to evaluate our type-based re-ranking algorithm using the Pliny and Grepcode search engines since they provide finer-grained support for querying methods.

\vspace{1em}
\noindent {\bf\emph{User study.}} Finally, a threat to the validity of our user study is the choice of programming tasks as well as the background of the participants. Specifically, 
only 62\% of the participants were experienced Java programmers, so it is conceivable that the participants in our study are not fully representative of professional Java developers. However, we believe that the potential users of a tool like \system also include novice programmers.


\section{Related Work}
In this section, we survey related work on code search, code reuse, and program synthesis.


\vspace{0.1in}
\noindent {\bf \emph{Code search and reuse.}} 
Existing techniques for code search and reuse can be divided into several categories, including 
textual~\cite{codeexchange,codebroker,sniff}, graph-based~\cite{sourcerer,portfolio,strathcona}, test-driven~\cite{codegenie, codeconjurer,s6,smt}, and type-based methods~\cite{parseweb,sigmatch1, sigmatch2}. 


\vspace{0.1in}
\noindent {\emph{Textual approaches}} 
 for code search identify relevant methods using keywords, such as comments and variable names~\cite{text1,text2,codebroker}. 
In more recent work, Chatterjee et al. have proposed annotating user code with documentations of API methods in order to improve free-form query search results~\cite{sniff}. Similarly, {Exemplar}~\cite{exemplar} uses help documents to expand queries and achieves multiple levels of granularity. CodeExchange~\cite{codeexchange} further refines textual methods by exploiting relationships between successive user queries. Unlike \system, none of these techniques enable \emph{fully automated reuse} of existing code.

\vspace{0.1in}
\noindent {\emph{Graph-based approaches}}  to code search represent relationships between software components as a graph and use a variant of the PageRank algorithm~\cite{pagerank} on the resulting graph representation. For example, Sourcerer~\cite{sourcerer} models software components as a directed graph, where nodes denote  classes and edges represent cross-component usage. This representation allows Sourcerer to  prioritize widely-used software components in the search results. Similarly, Portfolio~\cite{portfolio} generates a graph representation, where nodes correspond to methods and edges denote caller-callee relationships. The key insight underlying Portfolio is that methods that are called more frequently in the corpus are likely to be more relevant.

\vspace{0.1in}
\noindent {\emph{Type-based approaches}} allow users to formulate code search queries involving types. For example, Prospector~\cite{jungloid} and PARSEWeb~\cite{parseweb} can synthesize small code snippets that reuse existing functions based on queries of the form $(\tau_{in}, \tau_{out})$, where $\tau_{in}, \tau_{out}$ denote the desired input and output types respectively. Another related line of work is \emph{signature matching}, which aims to retrieve methods that (partially) match a type signature by reducing this problem to first-order unification ~\cite{sigmatch1,sigmatch2}. \system is closely related to these approaches in that it performs re-ranking and interface alignment based on type signatures. However, our approach differs from these techniques in that (i) we compute a type similarity metric based on multiset representations of classes, and (ii) solve an integer linear programming problem to find the best match. It is worth noting that \system uses an approach similar to Prospector and PARSEWeb for automatically synthesizing adapter code in a type-directed manner.

\vspace{0.1in}
\noindent {\emph{Test-driven approaches}} use test cases to partially specify the behavior of the desired method. For example, {CodeGenie}~\cite{codegenie} extracts the method signature and search keywords from JUnit tests provided by the user and queries {Sourcerer} to find relevant methods. It then uses the provided test cases to validate the search result and  merges the code into development environment. However, unlike \system, {CodeGenie} cannot synthesize adapter code.

Another test-driven tool  similar to \system is {CodeConjurer}~\cite{codeconjurer}, which allows users to specify  class components using test cases and UML-like interface description. Similar to \system, {CodeConjurer} can find a suitable mapping between the methods of the candidate class and those of the desired class. However, {CodeConjurer} tries all possible method-mapping permutations and cannot synthesize adapter code. Since the mapping technique is based on brute-force search, {CodeConjurer}'s method matching procedure can take several hours. In contrast, \system solves a different kind of interface alignment problem and finds an optimal solution through integer linear programming.

Among automated code reuse tools, \ssix~\cite{s6} is the most similar to \system. Specifically, \ssix uses a combination of test cases, method signatures, and natural language description to find relevant methods. Furthermore, similar to {\sc Hun-ter}, \ssix can adapt existing code to fit the desired interface by performing various kinds of transformations (e.g., parameter re-ordering and renaming). As demonstrated in our experiments, \system outperforms \ssix, both in terms of running time as well as the amount of code that it can reuse.

\vspace{0.1in}
\noindent {\bf \emph{Program synthesis and code completion.}} 
\system is also related to a long line of work on program synthesis and code completion. In particular, \system uses existing type-directed synthesis techniques to automatically generate any necessary adapter code.

In recent years, there has been a flurry of interest in automated code completion tools~\cite{codehint,insynth,jungloid,parseweb,xsnippet,strathcona,mapo,slang}. Some of these techniques use type information~\cite{insynth,jungloid,parseweb}, while others~\cite{mapo,slang} use statistical information about API usage patterns to find relevant completions. We believe that any type-directed  code completion tool can be gainfully integrated into \system for automatically synthesizing wrapper code from a candidate alignment.

Another related line of work is component-based synthesis~\cite{sypet,comp-pldi11,comp-icse10,comp-geometry}, which aims to synthesize programs from a database of underlying components. Component-based synthesis techniques have been  used in a  variety of application domains, including  bit-vector algorithms~\cite{comp-pldi11}, deobfuscators~\cite{comp-icse10},  and geometry constructions~\cite{comp-geometry}. Most recently, Feng et al. have proposed a type-directed approach to component-based synthesis that utilizes Petri net reachability analysis~\cite{sypet}. As mentioned earlier, \system internally uses Feng et al.'s {\sc SyPet} tool in order to synthesize adapter code in a type-directed manner.

\vspace{-0.05in}
\section{Conclusion}

We have presented a tool called \system for finding and automatically reusing existing methods. Given a search query and a set of test cases provided by the user, \system searches massive code repositories for candidate adaptee me\-thods and automatically synthesizes any necessary wrapper code. The key technical idea underlying \system is to compute a similarity metric between types and use this information for re-ranking search results and finding a suitable mapping between the parameters of the desired method and that of the adaptee. We formulate the problem of finding an optimal (i.e., lowest cost) mapping in terms of integer linear programming (ILP), and automatically synthesize wrapper code from the solution to the ILP problem  by leveraging existing type-directed synthesis techniques. Our experimental evaluation shows that \system compares favorably with existing code reuse tools and that our re-ranking algorithm improves the quality of the search results. We have also performed a user study involving 21 participants and 3 different programming tasks. Our study demonstrates that \system increases programmer productivity and helps programmers write less buggy code. The \system tool can be obtained for free from the Eclipse marketplace and used as an Eclipse plug-in to aid software development tasks.
\vspace{0.05in}


\balance
\bibliographystyle{abbrv}
\bibliography{main}

\end{document}